# Effect of metal dopant on structural and magnetic properties of ZnO nanoparticles

T. A. Abdel-Baset[1,2]*, M. Abdel-Hafiez[3]

[1] Department of Physics, Faculty of Science, Taibah University, Yanbu, 46423, Saudi Arabia.
tmohamed@taibahu.edu.sa

[2] Department of Physics, Faculty of Science, Fayoum University, Fayoum 63514, Egypt.
taa03@fayoum.edu.eg

[3] Department of Physics and Astronomy, Uppsala University, Uppsala 75120, Sweden,

**Abstract**

$Zn_{1-x}R_xO$ (R = Li, Mg, Cr, Mn, Fe and Cd) were obtained by using co-precipitation synthesis technique with constant weight percent of 3% from R ions. The phase composition, crystal structure, morphology, Density Functional Theory (DFT), and, magnetic properties were examined to comprehend the influence of $Zn^{2+}$ partial substitution with R ions. X-ray diffraction shows that the ZnO lattice parameters were slightly affected by R doping and the doped sample crystallinity is enhanced. Our results show that introducing Cr, Mn and Fe along with Mg into ZnO induces a clear magnetic moment without any apparent distortion in the structural morphology. The spatial configuration of dopant atoms is determined from first-principles calculations, giving a better understanding of the position of the dopant atom responsible for the magnetism. The magnetic moments obtained from our calculations are 3.67, 5.0, and 4.33 μB per dopant atom for Cr, Mn, and Fe, respectively, which agree with the experimental values. While Cr and Fe tend to form clusters, Mn has more propensity to remain evenly distributed within the system, avoiding cluster-derived magnetism.

## 1. Introduction

Magnetism in semiconductors has brought about the search for the next generation magnetic semiconductors. In these materials, rather than the electron charge, it is the electron spin that carries meaningful information [1]. These diluted magnetic semiconductors (DMSs) are formed when the cations in a non-magnetic semiconductor are partially replaced by magnetic transition metal ions [2-5]. They have garnered considerable interest as they possess the potential for interesting and practical applications in fields such as spintronics, spin-valve transistors, spin light-emitting diodes, and logic devices [6, 7]. One such DMS is the transition metal (TM) doped ZnO, which has attracted a lot of attention. This is due to its superior potential for applications in the field of spintronic devices [1-8]. Wurtzite ZnO has a direct band gap ($E_g$) of about 3.4 eV and high exciton binding energy of 60 meV. It shows piezoelectricity, and high thermal and chemical stability with respect to the environment [9-13]. The electron Hall mobility (~ 200 $cm^2$ $Vs^{-1}$) and Curie temperature for doped ZnO are also quite high [10-12]. These qualities render ZnO a material of paramount significance for a variety of optoelectronic and electronic applications [14]. Depending on the growth conditions, TM doping in the identical host lattice of ZnO can yield both paramagnetism as well as ferromagnetis [15-17]. ZnO in nano-scale can be synthesized by several different methods including physical vapor deposition [18], metal organic chemical vapor deposition [19], and a low temperature hydrothermal method [20]. The effect of doping on the structure, optical, electrical, and magnetic properties of ZnO have been studied by various groups [1,10].

The fundamental reason behind the origin of ferromagnetism in TM doped ZnO is still not fully understood. Different models have been put forward to explain the mechanism for the ferromagnetism at room temperature in DMSs. Some of these suggested mechansims on the nature of ferromagnetism are: (i) The interaction between bound magnetic polarons [21], (ii) a charge-transfer-based ferromagnetism model [22,23], (iii) oxygen and zinc vacancies, $V_O$ and $V_{Zn}$, interstitials, and (iv) O and Zn, surfaces and grain boundaries [24-30]. The possibility of ferromagnetism in Carbon-Doped ZnO [31] or nitrogen [32] has paved the way for further research. However, this type of magnetism is known as $d^0$ ferromagnetism [33] and experimentally observed in ZnO-based systems prepared by electron radiation [34] or ion implantation [35]. Zn interstitials or oxygen vacancies in C doped ZnO (ZnO:C) has also shown to enhance ferromagnetism in pulsed laser deposited films [36]. In general, the magnetic properties of TM-doped ZnO depends on the growth conditions. Particularly, these conditions form grain boundaries which resulted in ferromagnetic behavior. Importantly, the fundamental nature of the ferromagnetic behavior is debated and require further studies.

In this work, a detailed study on the preparation of high-quality samples of $Zn_{1-x}R_xO$ (R = Li, Mg, Cr, Mn, Fe and Cd) is reported, with constant weight percent of 3% from $R$ ions vs. ZnO. We found that the magnetic properties of ZnO strongly depended on type of R content in ZnO matrix. By controlling the $R$ content, different and unprecedented magnetic properties are reported, along with structural, morphological, and optical properties. For the *ab initio* calculations, we considered direct replacement of the Zn atoms by the dopants and did not consider Zn or O vacancies. While the Li and Cd doped systems remained paramagnetic, the Mg, Cr, Mn, and Fe-doped systems show ferromagnetism. In the Cr, Mn, and Fe-doped systems the presence of $d$ electrons can be held responsible for giving rise to magnetism, but a deeper explanation for the $d^0$ ferromagnetism in the Mg-doped sample provides a route for future work. Importantly, these properties can be taken advantage of in spintronic applications.

**2. Materials and Methods**

All the gradual steps leading to the preparation (weighing, mixing, grinding, and storage) were carried out in an Ar-filled glove box. Both $O_2$ and $H_2O$ levels were kept at less than 0.1 ppm. The preparation of $Zn_{1-x}R_xO$ ($x$ = 0.03) in a nanoparticles (NPs) form has been achieved by using the co-precipitation method. Zinc sulfate ($ZnSO_4$) and NaOH solutions were prepared separately and then eventually mixed together. This mixed solution was maintained at room temperature and stirred for 2 hours. For drying purposes, the $Zn(OH)_2$ was heated at 70 °C for 24 hours. The dried ingots were heated at 400 °C for 4 hours. Then the powder was left to cool down slowly to room temperature to get pure zinc oxide (ZnO). In order to prepare mixed oxide DMSs, mixed solutions of $ZnSO_4$ and the metal (R) sulfate were prepared. Finally, NaOH solution was added slowly to the mixed solution. This process was repeated to obtain $Zn_{0.97}R_{0.03}O$ nanoparticles for the different metals considered (R = Li, Mg, Cr, Mn, Fe and Cd).

All the structures including the pristine and doped ones were relaxed within the framework of Density Functional Theory (DFT) [37] by solving the Kohn-Sham equations[38]. Our total energy and electronic structure calculations were performed using the projector augmented wave (PAW) [39] formalism of density functional theory as implemented in the VASP package [40]. We used a well-converged energy cutoff of 550 eV and the Brillouin zone was sampled employing a 2 × 2 × 3 Monkhorst-Pack k-point mesh [41]. The Zn, O, Li, Mg, Cr, Mn, Fe and Cd potentials were employed with valence states of $3d^{10}4s^2$, $2s^22p^4$, $2s^1$, $3s^2$, $3p^63d^54s^1$, $3d^64s^1$, $3p^63d^54s^1$ and $4d^{10}5s^2$, utilizing the Perdew-Burke-Ernzerhof (PBE) exchange correlation functional [42]. All the structures were relaxed until the Hellmann-Feynman forces were less than 0.05 eV/Å. Typically, standard DFT calculations tend to grossly underestimate the electronic band gap of wide gap insulators or even semiconductors containing transition-metals.

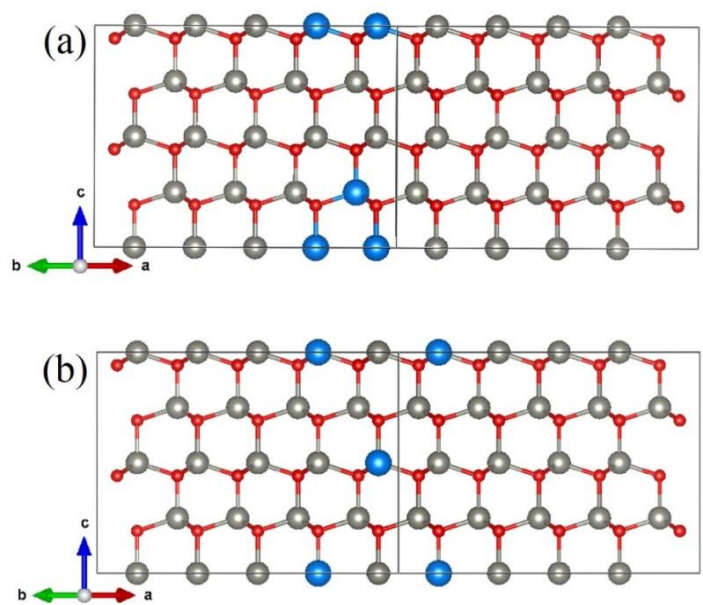

**Figure 1**. Wurtzite supercell of ZnO doped with 3% dopant atoms. The grey, red and blue atoms represents Zn, O and the dopant atoms respectively. Dopant atoms are separated by a) an oxygen atom (near), and b) –O-Zn-O- (far).

Therefore, a Hubbard parameter (U) is included to account for the *d-d* Coulomb interaction [43-45]. In order to add the on-site *d-d* Coulomb interaction, *U*, and the on-site exchange interaction, *J*, to the generalized gradient approximation (GGA) Hamiltonian for transition-metal elements, we use a typical value of U-J = 4.0 eV for our calculations that contain transition metals, *i.e.* Cr, Fe and Mn. In order to mimic the experimental doping concentration, we constructed a periodic 5×5×2 supercell of ZnO containing 200 atoms (100 Zn and 100 O atoms). Since, in the experiments, the doping concentration was 3%, in our calculations we replaced 3 of the Zn atoms with the dopant atoms.

In our calculations, we explored two different atomic positions of the dopants and checked for the minimum energy configuration. In the first configuration, the dopant atoms are separated by one oxygen atom which we designate as 'near' (Fig. 1a). In the other configuration, the dopant atoms are separated by –O-Zn-O- and we call this one the 'far' configuration (Fig. 1b). We then calculated the minimum energy configuration and did further analyses with those structures.

**3. Results and discussion**

*3.1. structure and morphology studies*

We first formed zinc Oxide NPs. In order to check the morphology, the average size, and the shape of the prepared nanoparticles, we have done both High resolution Transmission Electron Microscopy (HR-TEM) (Fig. 2a) and Scanning Electron Microscopy (SEM) (Fig. 2b). The synthesized NPs are generally isotropic in shape (Fig. 2a). Our ZnO NPs displaying the commonly observed wurtzite crystal structure (ICDD no. 36-1451, Fig. 2c) with a semi-spherical shape, and an average particle size is found to be about 14-16 nm, which is in good agreement with that obtained from TEM image and XRD analysis. It is well known that the cryzstallization of ZnO either presents a hexagonal wurtzite or cubic zincblende form. The latter is considered to be the most common. Our XRD confirms the synthesized ZnO NPs was free of impurities as it does not contain any characteristics XRD peaks other than zinc oxide peaks On the other hand, the surface properties of the grown NPs are of great interested. This is due to the molecules are arranged on the surface of the NPs since the surface to volume ratio for the NPs is higher than their bulk counterparts [10]. To check the existence of the any vibrational modes of ZnO NPs and to reveal the composition of our samples, FTIR Spectroscopy was measured in the range of 4000–400 cm$^{-1}$ is presented in Fig. 2d. The pronounced 450 cm$^{-1}$ and 3398 cm$^{-1}$ peak is the well known characteristic absorption of Zn-O bond and the characteristic absorption of hydroxy, repectively.

Under ambient conditions, ZnO has a hexagonal wurtzite structure with lattice parameters $a$ = 3.25 Å and $c$ = 5.20 Å [46]. In this work, we have, further, substituted 3% of the Zn atoms with Li, Mg, Cr, Fe, Mn and Cd. The crystal structures of the doped systems have been studied using X-ray diffraction (XRD) in order to find any deviation from that of pure ZnO. The XRD patterns of all the systems (pristine and doped) are shown in Fig. 3.

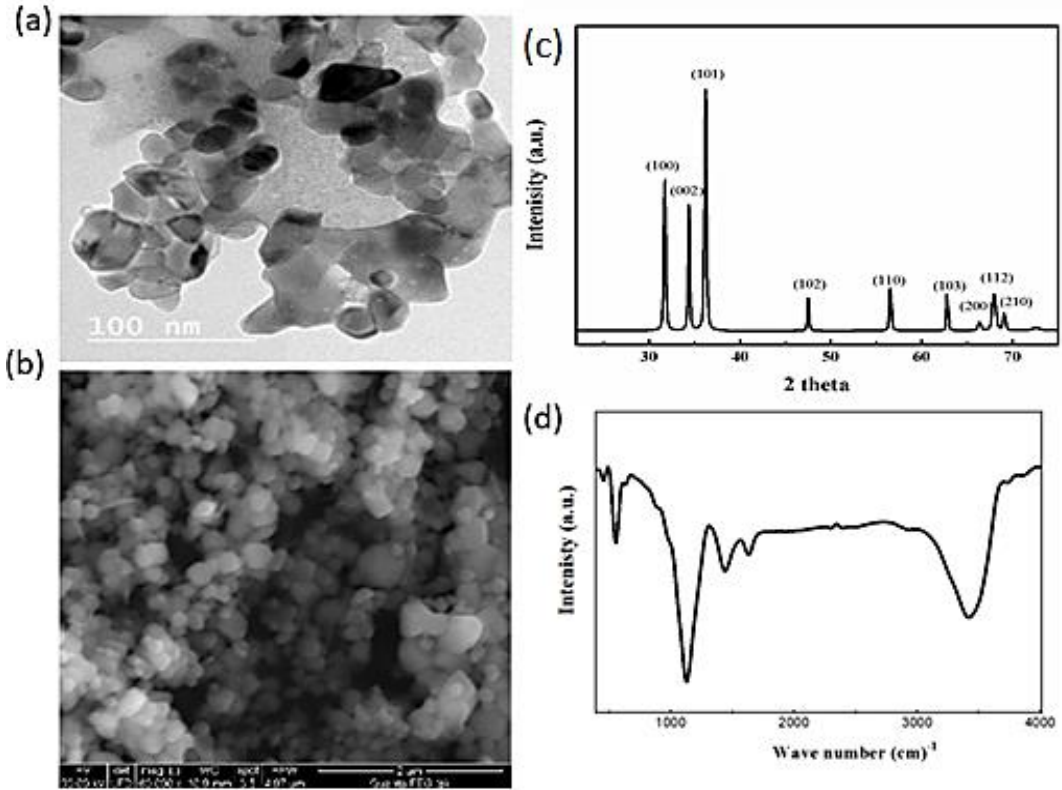

**Figure 2.** Characterization of typical ZnO Nanoparticles synthesized with the co-precipitation method. (a) TEM image (the scale bar is 100 nm). (b) SEM image. (c) XRD pattern (the ZnO wurtzite reference ICDD no. 36-1451 is highlighted). (d) Fourier-transform Infrared (FTIR) spectroscopy for ZnO Nanoparticles.

One can see that the X-ray diffraction patterns of the $Zn_{0.97}R_{0.03}O$ NPs compared to the pure ZnO nanostructure, where (R= Li, Mg, Cr, Mn, Fe and Cd). Pure ZnO has a hexagonal wurtzite structure with diffraction peaks (100), (002) (101), (102), (110), (103) and (112), which consistent with standard XRD data [see JCPDS 36-1451]. Only the characteristic peaks of ZnO were observed, which confirms that all synthesized samples are crystalline and are free of significant amounts of impurities. On other hand the peaks slightly shift with doping should be due to the different ionic radii of the dopant ions as follow, the peaks are shifted to higher angle for Fe (r = 0.77 Å), Mn (r = 0.80 Å) and Cd (r = 0.9 Å) doped ZnO, while for Cr (0.62 Å), Li (0.73 Å), and Mg (0.71 Å) is slightly shifted to lower angle doped ZnO, this shifted may be due to the different ionic radii of the dopant ions with the ionic radius of Zn (0.74 Å). Furthermore, with the addition of *R* ions into ZnO, the intensity of diffraction peaks slightly increased whereas after insertion of Cr ions into ZnO matrix the diffraction peaks for $Zn_{0.97}Cr_{0.03}O$ becomes more intense. This may be due to the improvement in the crystallinity of the formed composite. The lattice parameters *a* and *c* are determined from the following relation,

$$\frac{1}{d^2_{(hkl)}} = \frac{4}{3}\left[\frac{h^2+hk+k^2}{a^2}\right] + \frac{l^2}{c^2} \qquad (1)$$

where *h*, *k* and *l* are the Miller Indices, *d* is interplanar distance and '*a*' and '*c*' are lattice parameters, and listed in Table (1). Average crystallite sizes (D) were calculated by using the Debye-Scherrer's equation ($D = k\lambda/\beta \cos\theta$), and are also listed in Table (1), where $k$ = 0.94 (a constant), $\lambda$ is wavelength of the X-rays, $\beta$ is full width at half maximum (FWHM) and θ is the diffraction angle.

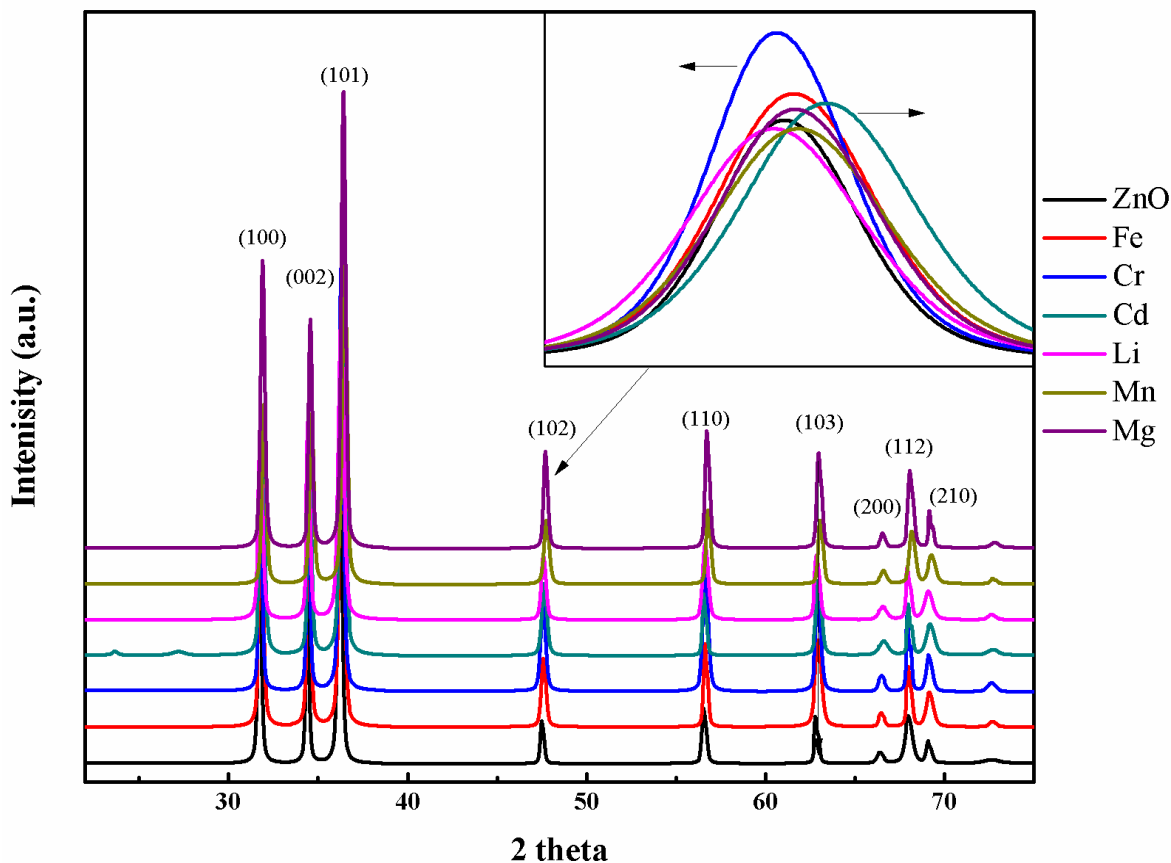

**Figure 3.** XRD patterns of $Zn_{1-x}R_xO$ (R= Li, Mg, Cr, Mn, Fe and Cd).

Table 1. Crystallite size, lattice strain, lattice parameters and the unit cell volume (v) of pure ZnO and $Zn_{0.97}R_{0.03}O$ nanoparticles.

| $Zn_{0.97}R_{0.03}O$ | $(D)$(nm) | Lattice Strain (ε) % | $a$(Å) | $c$(Å) | $c/a$ | $V$(Å$^3$) |
|---|---|---|---|---|---|---|
| Pure ZnO | 14.25 | 0.241 | 3.2487 | 5.2033 | 1.6017 | 54.92 |
| R= Fe | 16.58 | 0.266 | 3.251 | 5.209 | 1.6023 | 55.05 |
| Cr | 13.98 | 0.225 | 3.2485 | 5.21 | 1.6038 | 54.98 |
| Cd | 18.68 | 0.385 | 3.25 | 5.201 | 1.6003 | 54.94 |
| Li | 14.08 | 0.405 | 3.254 | 5.215 | 1.6026 | 55.22 |
| Mn | 17.85 | 0.351 | 3.2504 | 5.2057 | 1.6016 | 55 |
| Mg | 14.17 | 0.352 | 3.2501 | 5.2113 | 1.6034 | 55.05 |

One can see that the Li and Mg doped sampleshave smaller average crystallite sizes, compared to the undoped ZnO. The ionic radii of $Zn^{2+}$ is 0.74 Å while that of $Li^{2+}$ and $Mg^{2+}$ are 0.73 Å and 0.71 Å, respectively, for four-fold coordination. Zn in ZnO is tetrahedraly coordinated with oxygen. Therefore, the decrement in the crystal size for the Li and Mg substituted ZnO in place of Zn is expected. Chromium in the four-fold coordination can have valencies 4+, 5+ and 6+ and their ionic radii are 25.6%, 34.5% and 45.9% smaller than that of $Zn^{2+}$. Therefore, the smaller crystal size of

$Zn_{0.97}Cr_{0.03}O$ compared to pure ZnO is also justified. In order to maintain charge neutrality, Fe and Mn should have valency of 2+. But the ionic radii of $Fe^{2+}$, $Cd^{2+}$ and $Mn^{2+}$ are 0.77 Å, 0.90 Å and 0.80 Å, respectively. Therefore, on substituting $Zn^{2+}$ with these ions the crystallite size is found to be increased may be due to an increase in nucleation rates of ZnO nanoparticles Also, the slight variation in lattice parameter likely results from the substitution of different ionic radii doped ions with Zn. Furthermore, it is noted that there is an increase in its lattice strain ($\varepsilon = \beta_{hkl}\cos\theta/4$) due to R-doping inside the ZnO matrix (Table 1), which causes the local distortion of the crystal structure. This is expected and has been found in previous studies [47].

In order to further check both the average size and shape of our grown samples, we have done SEM for all grown NPs (Fig. 4). The morphology of $Zn_{0.97}R_{0.03}O$ nanoparticles were also studied using SEM imaging (Fig. 4: a-f). The SEM image represents the agglomeration of particles and also shows a narrow particle size distribution.

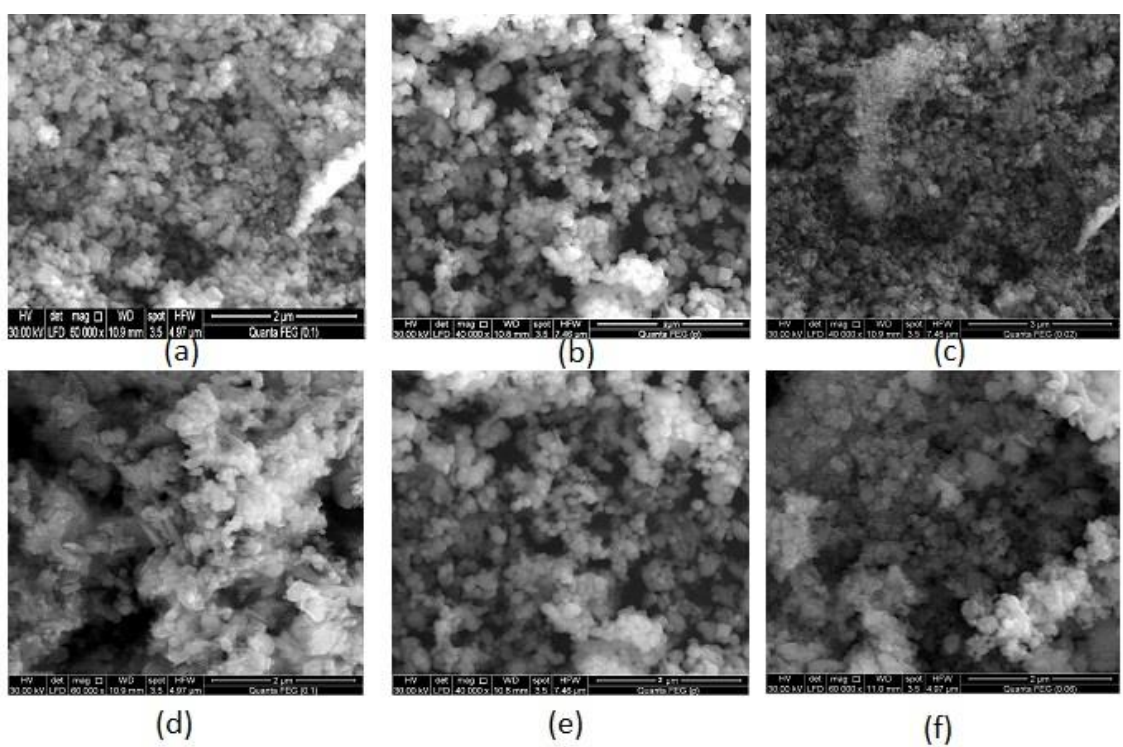

**Figure** 4 (a-f). SEM images for $Zn_{1-x}R_xO$ samples, for 3% R concentrations a. Mn, b. Cr, c. Cd, d. Li, e. Mg, and f. Fe, respectively.

We have, further, performed Fourier-transform Infrared (FTIR) spectroscopy for all investigated samples to determine the vibrational modes present. Figure 5 shows the FTIR spectrum of the $Zn_{0.97}R_{0.03}O$ nanoparticles compared to the pure ZnO, where (R = Li, Mg, Cr, Mn, Fe and Cd) synthesized from zinc sulfate acquired in the range of 400-4000 $cm^{-1}$ observed at different regions of FTIR spectrum. The observed spectra exhibited well defined absorption peaks in the range of 550 $cm^{-1}$ which is a manifestation of the stretching vibrational mode of ZnO. These data are similar to the results observed perviously [48].

The absorption peaks in the range of 460-550 $cm^{-1}$ which can be attributed to the stretching vibrational mode of ZnO [48]. In addition, the observed strong band at 459 $cm^{-1}$ indicates the existence of pure ZnO which shifted to 463$cm^{-1}$ as a result of metal dopant ions. Shoulder peaks are seen to exist around the bands at around 1420 cm$^{-1}$. The peak around 1636 cm$^{-1}$ is due to the H–O–H bending vibration. The absorption peak that appears at 3420 cm-1 is due to the O–H stretching vibrations of H$_2$O, which is present in trace amounts of the nanomaterial [49]. Due to the replacement of R for Zn, the Zn0.97R0.03O bands display a small difference in peak position compared to the pure ZnO sample. The same results were reported for Ni and Mn doped ZnO NPs [1,2].

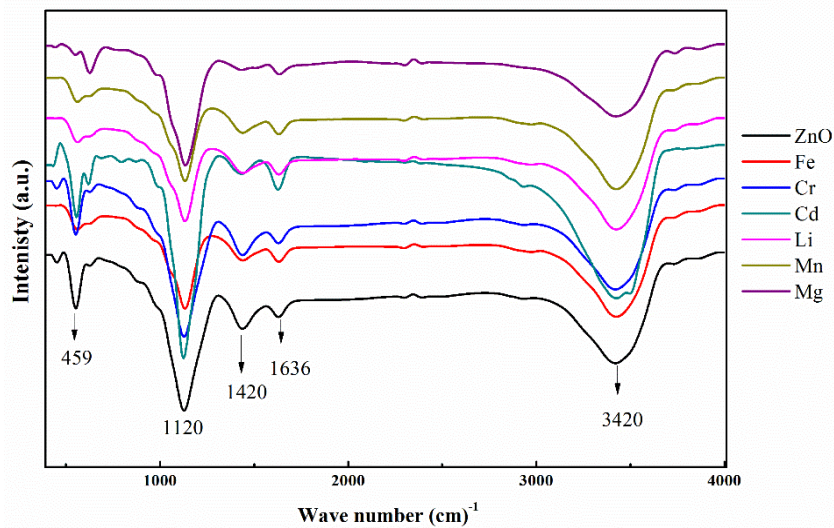

**Figure** 5. FTIR spectra of pure ZnO and Zn$_{0.97}$R$_{0.03}$O samples.

*3.2. DFT Calculations*

The energy difference of the two configurations is shown in Fig. 6, left panel. It is clear that for Li, Mg, and Cd, the far configuration is energetically more favorable. Although, for Mg and Cd, the energy differences between the near and far configurations are very small compared to that of Li, we still proceeded with the far configuration for our calculations. For the transition metals, Cr and Fe, the near configuration turns out to be the lower energy structure, while it is opposite for Mn. This trend for Cr, Mn and Fe is in excellent agreement with that obtained by Gopal *et al.* [50].

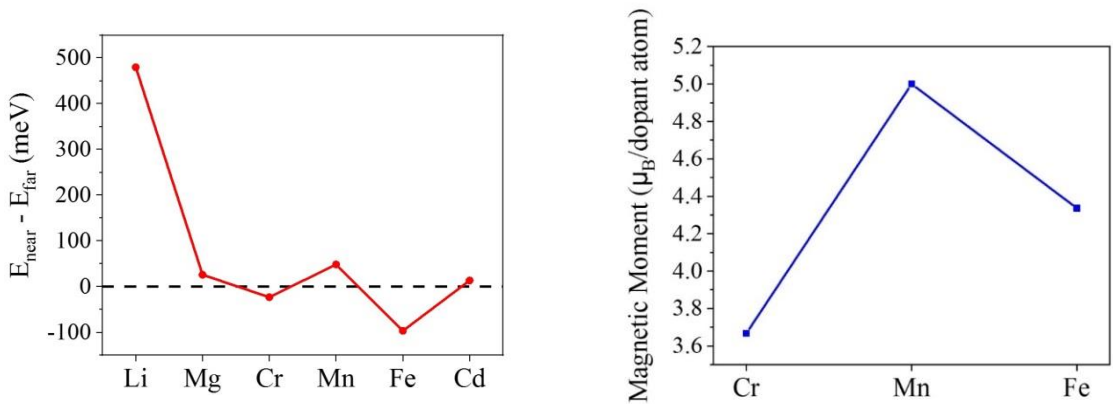

**Figure** 6. left panel: $E_{near}$ - $E_{far}$ for substitutional dopant ions. A negative energy difference indicates that the TM ions prefer to be in a near spatial configuration. Right panel: Magnetic moment per dopant atom

The magnitudes here are different from those shown in Ref. 50. This is due to the difference in the dopant concentration and the different pseudopotentials used in our calculations. The doped structures were allowed to relax with respect to all the degrees of freedom and eventually, with the end results showing agreement with the experimental findings. For all the doped cases, the wurtzite morphology of ZnO is maintained and no structural phase transition is observed, as seen in our XRD measurements. The structural parameters of all the doped systems is presented in Table 2. The lattice parameters don't show significant differences for the different dopant cases in comparison to the pristine wurtzite ZnO structure. The maximum deviation is seen for Cd, which can be attributed to the larger ionic radii. The almost constant c/a value confirms the consistency of the wurtzite structure. As can be seen, the experimental (Table 1) and calculated (Table 2) data are in very close agreement.

Table 2. Lattice constants and the dopant-oxygen bond length obtained from calculations.

| Dopant (3 % doping) | Avg. M-O bond length (Å) | a (Å) | c (Å) | c/a |
|---|---|---|---|---|
| Li | 1.965 (8) | 3.2488 (1) | 5.2240 (1) | 1.6079 (7) |
| Mg | 1.960 (6) | 3.2557 (1) | 5.225 (0) | 1.6049 (0) |
| Cr | 2.040 (5) | 3.2597 (4) | 5.2331 (4) | 1.6053 (8) |
| Mn | 2.012 (1) | 3.2588 (6) | 3.2355 (3) | 1.6065 (5) |
| Fe | 1.971 (7) | 3.2571 (7) | 5.2352 (4) | 1.6072 (9) |
| Cd | 2.169 (2) | 3.2635 (6) | 5.2449 (0) | 1.6071 (1) |

With respect to chemical bonding, ZnO is a tricky material as it shows a mixture of both ionic and covalent bonding. The electron transfer from Zn to O is 1.15, which is not strong enough for ionic bonding or not enough for covalent bonding. We calculated the Bader charge transfer (Table 3) using the Atoms in Molecules Theory [51]. Amongst, all the doped systems studied in this work, the Zn-O bonds still remain partially covalent and partially ionic. However, for Li, Mg, Cr, Mn and Fe, the dopant and oxygen bonding tend more towards being ionic. Given that Zn and Cd belong to the same group elements ($d^{10}$), the electron transfer from them to oxygen is almost the same, and therefore the Cd-doped system retains both the ionic and covalent traits.

Table 3. Bader charge transfer from Zn and dopant atoms to O

| Dopant (3 % doping) | Number of Electrons Donated by Zn to O | Number of Electrons Donated by Dopant to O |
|---|---|---|
| Li | 1.15 | 0.84 |
| Mg | 1.15 | 1.63 |
| Cr | 1.15 | 1.35 |
| Mn | 1.15 | 1.33 |
| Fe | 1.15 | 1.31 |
| Cd | 1.15 | 1.13 |

Experimentally (*vide infra*), we observed that by introducing Cr, Mn, and Fe to the ZnO, clear hysteresis loops can be observed and the existence of ferromagnetism in these materials has been verified. From our calculations, the magnetic moments that were obtained are 3.67, 5.0, and 4.33 $\mu_B$ per dopant atom for Cr, Mn, and Fe respectively (Fig. 6, Right panel) [52,53]. While Cr and Fe tend to form clusters, Mn has more propensity to remain evenly distributed within the system, avoiding cluster-derived magnetism. As reported by Xing *et al.*, the occurrence of ferromagnetism in transition metal-doped ZnO could possibly be due to the increase in the number of defects and oxygen vacancies [54]. Another opinion, according Chu *et al.*, is that the exchange interaction between transition metal ions and O ion spin moments can lead to ferromagnetism [55].

*3.3. Magnetic Proeprties*

The temperature dependencies of the magnetization upon zero-field cooling (ZFC) and field cooling (FC) are shown in Fig. 8 for the doped ZnO samples. A distinct ferromagnetic transition is seen for the Fe/Mn/Cr/Mg-doped samples. The ferromagnetic transition temperature ($T_c$) is estimated from the derivative of the magnetization data. Iron doping enhances the ferromagnetism of $Zn_{1-x}Fe_xO$ with the $T_c$ increasing from 39 K for 2% doping to 44 K for 8% doped samples [52]. As shown in [52] the decrement vanishes upon applying high magnetic fields and the ZFC curve overlaps with the FC curve. This downturn is similar to that seen in Ref. 43 and can be attributed to the antiphase ferromagnetic domains.

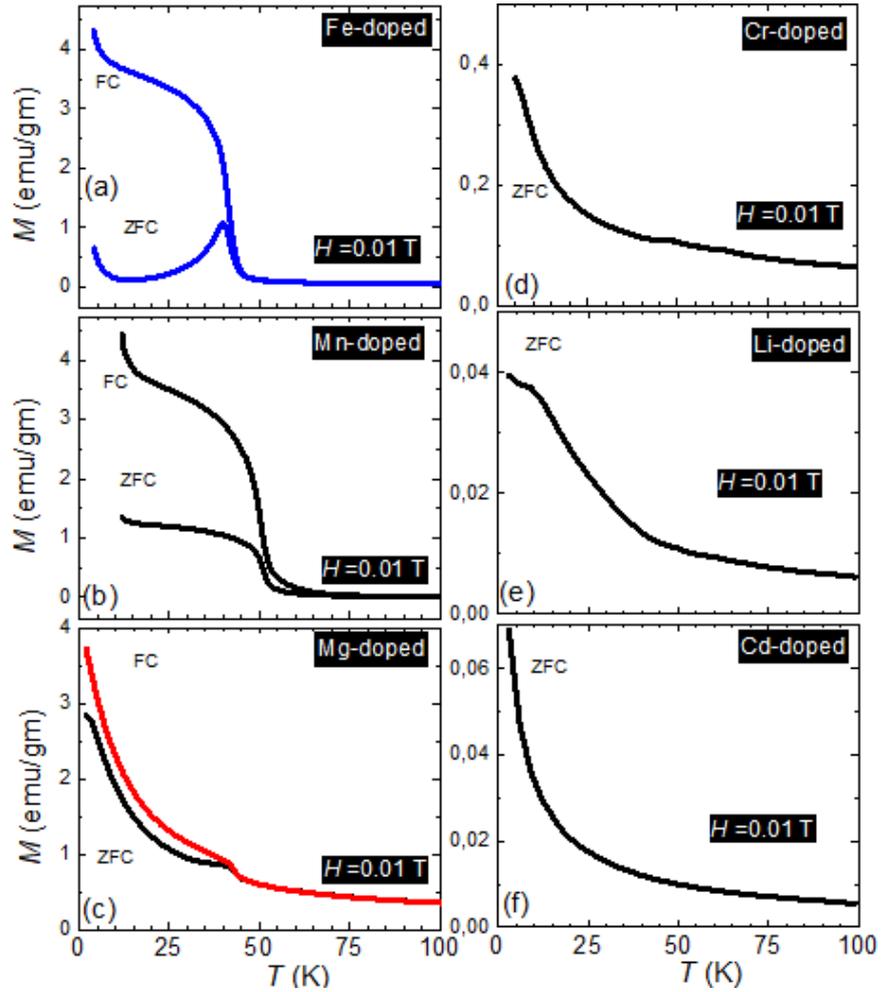

**Figure** 8. The temperature dependence of the magnetization (*M*) in an external field of 0.01T. *M* has been deduced from the dc magnetization measured by following ZFC and FC protocols.

One can noticed that the doped samples exhibit a clear magnetic hysteresis loop (Fig. 9) for the Mg, Cr, Mn, and Fe cases. Substituting Mg, Cr, Mn, and Fe ions into ZnO induced weak ferromagnetism and illustrates a magnetic moment without any distortion in the geometrical symmetry. It is important to note that the reduced moment values might occur due to large numbers of disordered spins on the surface. It could also be due to grain boundaries that constitute a considerable fraction of volume of the nano-sample. Vacancies will be ordered along the unique axis of magnetization (c-axis) in our case, as the samples exhibit hexagonal symmetry.

Importantly, it is well known that undoped ZnO is diamagnetic in nature. Thus, the ferromagnetism observed in our investigated samples is intrinsic and can not be attributed to any kind of defect [56]. Therefore, ferromagnetism in our investigated samples could be arising from spin-orbit interaction, through *s-d* orbitals coupling [57], or interaction between bound magnetic polarons. In the latter case, the fluctuations of magnetization have an important role in binding and may cause non-trivial thermal behavior [58]. Thus, the formation of charge carrier quantum states exists on the doped ZnO. However, the interaction between bound magnetic polarons is still not well understood when considering its role in ferromagnetism.

As inferred by Karmakar *et al*. [59], native defects in the ZnO host can give rise to ferromagnetism in the TM-doped ZnO system. According to Jin *et al*. [60], the bound magnetic polaron (BMP) model can be applied to explain room temperature ferromagnetism. The BMP model puts forward the idea that the long-range ordering is either due to the direct overlaps between BMPs or indirect interactions between the BMPs and magnetic impurities. Thus, doping with Cr increases the density of magnetic impurities, facilitating the ferromagnetic alignment of BMPs. It has also been proposed in the literature [61] that the spin-polarized carriers and the hybridization between the O-p and TM-d orbitals may be responsible for the ferromagnetism in Cr-doped ZnO. From our calculations the average magnetic moment of each Cr atom is found to be about 3.65 µB/Cr. This value is much larger than our experimental data. The reason behind this discrepancy can be explained as follows. The processes of preparing a film is often a non-equilibrium process. During the process, Cr atoms may get embedded in the antiferromagnetic far state which does not contribute to the ferromagnetism. It is widely proposed that defects, Zn or O vacancies, along with the presence of magnetic dopants, also play a crucial role in the enhancement or suppression of ferromagnetism [62-69]. However, in our calculations, the simplest cases have been treated with only replacement of Zn atoms by the dopants.

In the absence of magnetic impurities, exchange interactions between localized electron spin moments resulting from oxygen vacancies at the surfaces of nanoparticles can be the cause of ferromagnetism. A common consensus with regards to the ferromagnetic property in transition-metal-doped ZnO is that defects such as oxygen or zinc vacancies might be responsible [70-72]. It has already been shown that oxygen vacancy is crucial for Mn doped $In_2O_3$ [73]. Although the reason behind the intrinsic ferromagnetism in Mn-doped oxides is still not completely understood, it is believed that defects do play a role in the origin of the ferromagnetism in the Mn-doped ZnO system. The theoretical work by Gopal *et al.* shows that in TM-doped ZnO, and particularly Co-doped ZnO, the defects do favor ferromagnetic states [41]. This makes the claim for Mn-doped ZnO stronger. Finally, the diamagnetic dilution in the hexagonal diamagnetic lattice films of wurtzite structure resulted in improved values of the magnetic moment and ferromagnetic character which need further investigation. We believe that, for the case of Mg in our work, it is defects such as Zn and O vacancies which give rise to the $d^0$ ferromagnetism. From our theoretical calculations, we did not consider any defects ($V_{Zn}$ or $O_{Zn}$) in the lattice and also did not find any ferromagnetic moment for the Mg doped case. In general, the occurrence and improvement in the ferromagnetic behavior is also dependent on the synthesis route.

Fig.9: (a-d) show the isothermal magnetization *M* vs. *H* loops of $Zn_{0.97}R_{0.03}O$ [R=Fe (a), Mn(b), Mg(c), Cr(d)] measured at different temperatures in (a) ranging from 1.8 to 300K up to 7T. From (b-d) the isothermal magnetization M vs. H being collected at 300K.

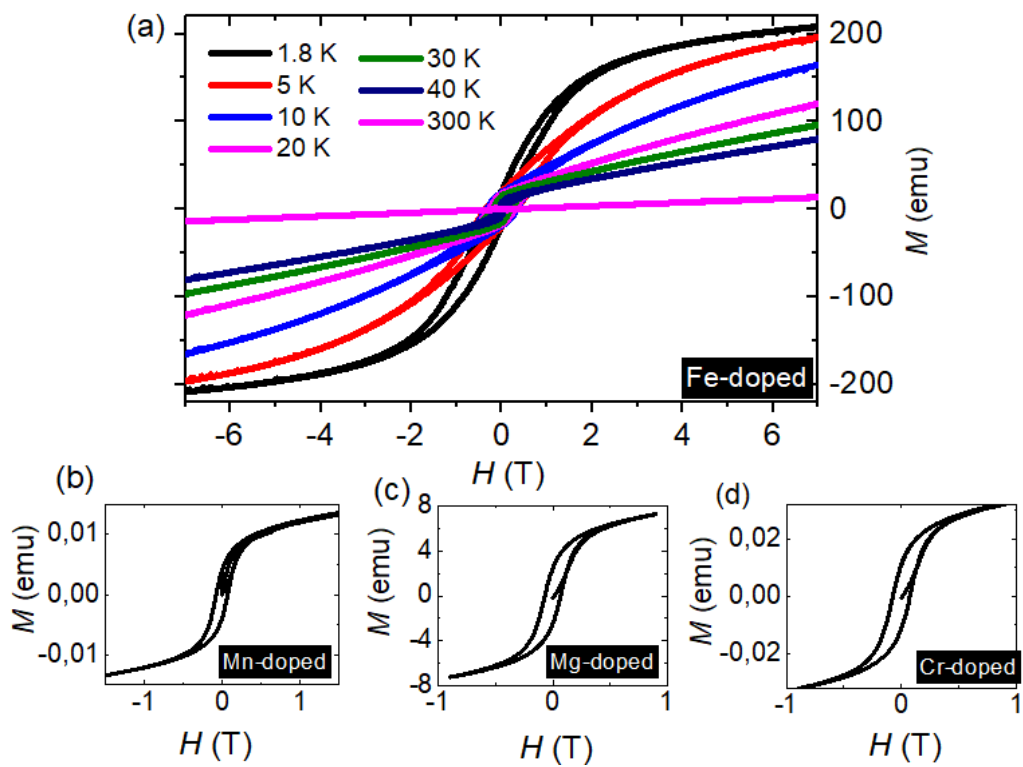

**Figure** 9: (a-d) show the isothermal magnetization *M* vs. *H* loops of $Zn_{0.97}R_{0.03}O$ [R=Fe (a), Mn(b), Mg(c), Cr(d)] measured at different temperatures in (a) ranging from 1.8 to 300K up to 7T. From (b-d) the isothermal magnetization *M* vs. *H* being collected at 300K.

**4. Conclusions**

In summary, pure and metal (Li, Mg, Cr, Mn, Fe and Cd)-doped ZnO nanoparticles were successfully synthesized. The original hexagonal wurtzite structure (*P6₃mc*) was retained for all samples. Pure ZnO exhibits paramagnetic behavior, as expected, but the Fe, Mn, Mg, and Cr doping induced ferromagnetism at different levels. The spatial configuration of the dopant atoms is determined theoretically, giving a better understanding of the dopant atom orientation responsible for magnetism. The Fe and Cr atoms tend to form clusters giving rise to ferromagnetism, while the Mn atoms remain more spread out. Several reasons behind the observed ferromagnetism have been discussed. Specifically, the role of defects has been emphasized when explaining the observed ferromagnetism. This is further strengthened by the case of the Mg (non-magnetic) doped case, where the formation of defects in the system can give the best justification for the occurrence of ferromagnetism.

**References**

1. T. Dietl, H. Ohno, F. Matsukara, J. Cibert & D. Ferrand, Zener Model Description of Ferromagnetism in Zinc-Blende, *Science*., 2000, **287,** 1019–1023.
2. A. G. EL. Hachimi, H. Zaari, A. Benyoussef, M. El. Yadari & A. El. Kenz, First-principles prediction of the magnetism of 4f rare-earth-metal-doped wurtzite zinc oxide, *J. Rare Earths.*, 2014, **32,** 715–721.

3. M. Zhang *et al*, Transition Metal Adsorbed-Doped ZnO Monolayer : 2D Dilute Magnetic Semiconductor , Magnetic Mechanism , and Beyond 2D, *ACS Omega.*, 2017, **2,** 1192–1197.
4. S. A. Ahmed, Room-temperature ferromagnetism in Co- , Cr- , and V-doped ZnO diluted magnetic semiconductor, *Appl. Phys. A.*, 2017, **123,** 440.
5. Z. Jin *et al*, High throughput fabrication of transition- metal-doped epitaxial ZnO thin films : A series of oxide-diluted magnetic semiconductors and their properties, *Appl. Phys. Lett.*, 2001, **78,** 3824–3826.
6. J. H. Zheng, J. L. Song, Z. Zhao, Q. Jiang & J.S. Lian, Optical and magnetic properties of Nd-doped ZnO nanoparticles, *Cryst. Res. Technol.*, 2012, **47,** 713–718.
7. O. Mounkachi *et al*, Magnetic properties of Zn0.9(Mn0.05,Ni0.05)O nanoparticle: Experimental and theoretical investigation, *J. Magn. Magn. Mater.*, 2012, **324,** 1945–1947.
8. O. Mounkachi, A. Benyoussef, A. El. Kenz, E. H. Saidi & E. K. Hill, Electronic structure of acceptor defects in ( Zn , Mn ) O and ( Zn , Mn )( O , N ), *J. Appl. Phys.*, 2009, **106,** 093905.
9. H. Gong, J.Q. Hu, J H. Wang, C. H. Ong, & F. R. Zhu, Nano-crystalline Cu-doped ZnO thin film gas sensor for CO, *Sensors Actuators B.*, 2006, **115,** 247–251.
10. S. Yodyingyong *et al*, ZnO nanoparticles and nanowire array hybrid photoanodes for dye-sensitized solar cells, *Appl. Phys. Lett.*, 2010, **96,** 073115.
11. M. J. Zheng, L. D. Zhang, G. H. Li, & W. Z. Shen, Fabrication and optical properties of large-scale uniform zinc oxide nanowire arrays by one-step electrochemical deposition technique, *Chem. Phys. Lett.*, 2002, **363,** 123–128.
12. L. K. Adams, D. Y. Lyon & P. J. J. Alvarez, Comparative eco-toxicity of nanoscale $TiO_2$ , $SiO_2$ , and ZnO water suspensions, *Water Res.*, 2006, **40,** 3527–3532.
13. P. Nunes *et al*, Effect of different dopant elements on the properties of ZnO thin films, *Vacuum.*, 2002, **64,** 281–285.
14. A. Panwar & K. L. Yadav, A novel one-pot synthesis of hierarchical europium doped ZnO nanoflowers. *Mater. Lett.*, 2015, **142,** 30–34.
15. D. More *et al*, Correlation of structural and magnetic properties of Ni-doped ZnO nanocrystals, *J. Phys. D. Appl. Phys.*, 2014, **47,** 045308.
16. S. J. Pearton, W. H. Heo, M. Ivill, D. P. Norton & T. Steiner, Dilute magnetic semiconducting oxides, *Semicond. Sci. Technol. Top.*, 2004, **19,** R59–R74.
17. I. Zutic, J. Fabian & S. D, Sarma, Spintronics: Fundamentals and applications, *Rev. Mod. Phys.*, 2004, **76,** 323–410.
18. Q. Li *et al*, Photoluminescence and wetting behavior of ZnO nanoparticles / nanorods array synthesized by thermal evaporation, *J. Alloys Compd.*, 2013, **560,** 156–160.
19. P. Hu, N. Han, D, Zhang, J. C. Ho, & Y. Chen, Chemical Highly formaldehyde-sensitive, transition-metal doped ZnO nanorods prepared by plasma-enhanced chemical vapor deposition, *Sensors Actuators B. Chem.*, 2012, **169,** 74–80.
20. R. Shi, P. Yang, X. Dong, Q. Ma & A. Zhang, Growth of flower-like ZnO on ZnO nanorod arrays created on zinc substrate through low-temperature hydrothermal synthesis, *Appl. Surf. Sci.*, 2013, **264,** 162–170.
21. J. M . D. Coey, M. Venkatesan & C. B. Fitzgerald, Donor impurity band exchange in dilute ferromagnetic oxides, *Nat. Mater.*, 2005, **4,** 173–179.
22. J. M. D. Coey, K. Wongsaprom, J. Alaria, & M. Venkatesan, Charge-transfer ferromagnetism in oxide nanoparticles, *J. Phys. D. Appl. Phys.*, 2008, **41,** 134012.

23. J. M . D. Coey, P. Stamenov, R, D. Gunning, M. Venkatesan & K. Paul, K, Ferromagnetism in defect-ridden oxides and related materials, *New J. Phys.,* 2010, **12,** 053025.
24. D. E. Motaung *et al*, Shape-Selective Dependence of Room Temperature Ferromagnetism Induced by Hierarchical ZnO Nanostructures, *ACS Appl. Mater. Interfaces.,* 2014, **6,** 8981−8995.
25. J. J. Beltran, C. A. Barrero & A. Punnoose, Understanding the role of iron in the magnetism of Fe doped ZnO nanoparticles, *Phys. Chem. Chem. Phys.,* 2015, **17,** 15284–15296.
26. X. Xu *et al*, Size Dependence of Defect-Induced Room Temperature Ferromagnetism in Undoped ZnO Nanoparticles, *J. Phys. Chem. C.,* 2012, **116,** 8813−8818.
27. N. Sanchez, S. Gallego, J. Cerdá & M. C. Muñoz, Tuning surface metallicity and ferromagnetism by hydrogen adsorption at the polar ZnO (0001) surface, *Phys. Rev. B.,* 2010, **81,** 115301.
28. A. L. Schoenhalz, J. T. Arantes, A. Fazzio & G. M. Dalpian, Surface magnetization in non-doped ZnO nanostructures, *Appl. Phys. Lett.,* 2009, **94,** 162503.
29. P. Zhan *et al*, Origin of the defects-induced ferromagnetism in un-doped ZnO single crystals, *Appl. Phys. Lett.,* 2013, **102,** 071914.
30. S. Lany, J. Osorio-guillén & A. Zunger, Origins of the doping asymmetry in oxides: Hole doping in NiO versus electron doping in ZnO, *Phys. Rev. B.,* 2007, **75,** 241203.
31. H. Pan *et al*, Room-Temperature Ferromagnetism in Carbon-Doped ZnO, *Phys. Rev. Lett.,* 2007, **99,** 127201.
32. I. S. Elfimov *et al*, Magnetizing Oxides by Substituting Nitrogen for Oxygen, *Phys. Rev. Lett.,* 2007, **98,** 137202.
33. J. M. D. Coey, d0 ferromagnetism, *Solid State Sci.,* 2005, **7,** 660–667.
34. L. S. Vlasenko & G. D. Watkins, Optical detection of electron paramagnetic resonance for intrinsic defects produced in ZnO by 2.5-MeV electron irradiation in situ at 4.2 K, *Phys. Rev. B.,* 2005, **72,** 035203.
35. S. O. Kucheyev, J. S. Williams & C. Jagadish, Ion-beam-defect processes in group-III nitrides and ZnO, *Vacuum.,* 2004, **73,** 93–104.
36. X. Li, J. Guo, Z. Quan, X. Xu & G. A. Gehring, Defects Inducing Ferromagnetism in Carbon-Doped ZnO Films, *IEEE Trans. Magn.,* 2010, **46,** 1382–1384.
37. P. Hohenberg & W. Kohn, Inhomogeneous Electron Gas, *Phys. Rev.,* 1964, **136,** B864.
38. W. Kohn & L. J. Sham, Self-Consistent Equations Including Exchange and Correlation Effects, *Phys. Rev.,* 1965, **140,** A1133.
39. P. E. Blochl, Projector-Augmented Plane-Wave Method, *Phys. Rev. B.,* 1994, **50,** 17953–18979.
40. G. Kresse & J. Hafner, Ab initio molecular dynamics for liquid metlas, *Phys. Rev. B.,* 1993, **47,** 558–561.
41. H. J. Monkhorst & J. D. Pack, Special points for Brillonin-zone integrations, *Phys. Rev. B.,* 1976, **13,** 5188–5192.
42. J. P. Perdew, K. Burke & M. Ernzerhof, Generalized Gradient Approximation Made Simple, *Phys. Rev. Lett.,* 1996, **77,** 3865–3868.
43. V. I. Anisimov, J. Zaa nen & O. K. Andersen, Band theory and Mott insulators: Hubbard U instead of Stoner I, *Phys. Rev. B.,* 1991, **44,** 943–954.
44. W. E. Pickett, S. C. Erwin & E. C. Ethridge, Reformulation of the LDA ⁄ U method for a local-orbital basis, *Phys. Rev. B.,* 1998, **58,** 1201–1209.
45. M. Cococcioni & S. De. Gironcoli, Linear response approach to the calculation of the effective

interaction parameters in the LDA+ U method, *Phys. Rev. B.,* 2005, **71,** 035105.

46. J. Albertsson, S. C. Abrahams & A. Kvick, Atomic Displacement, Anharmonie Thermal Vibration, Expansivity and Pyroeleetrie Coefficient Thermal Dependences in ZnO, *Acta Crystallogr.* B**.,** 1989, **45,** 34–40.

74. K. Irshad, M. Tahir & A. Murtaza, Synthesis and characterization of transition-metals-doped ZnO nanoparticles by sol-gel auto-combustion method, *Phys. B Phys. Condens. Matter.,* 2018, **543,** 1–6.

48. S. Bhatia, N. Verma &R. K. Bedi, Ethanol gas sensor based upon ZnO nanoparticles prepared by different techniques, *Results Phys.,* 2017, **7,** 801–806.

49. N. V. Kaneva & C. D. Dushkin, Preparation of nanocrystalline thin films of ZnO by sol-gel dip coating, *Bulg. Chem. Commun.,* 2011, **43,** 259–263.

50. P. Gopal & N. A. Spaldin, Magnetic interactions in transition-metal-doped ZnO : An ab initio study, *Phys. Rev. B.,* 2006, **74,** 094418.

51. G. Henkelman, A. Arnaldsson & H. A. Jonsson, A fast and robust algorithm for Bader decomposition of charge density, *Comput. Mater. Sci.,* 2006, **36,** 354–360.

52. T. A. Abdel-Baset, Y. Fang, B. Anis, C. Duan & M. Abdel-hafiez, Structural and Magnetic Properties of Transition-Metal-Doped $Zn_{1-x}Fe_xO$. *Nanoscale Res. Lett.* 2016, **11,** 115.

53. T. A. Abdel-Baset, Y. Fang, C. Duan & M. Abdel-hafiez, Magnetic Properties of Chromium-Doped ZnO, *J. Supercond. Nov. Magn.,* 2016, **29,** 1937–1942.

54. G. Z. Xing *et al,* Strong correlation between ferromagnetism and oxygen deficiency in Cr-doped $In_2O_{3-\delta}$ nanostructures, *Phys. Rev. B.,* 2009, **79,** 174406.

55. D. Chu, Y. Zeng & D. Jiang, Synthesis and growth mechanism of Cr-doped ZnO single-crystalline nanowires, *Solid State Commun.,* 2007, **143,** 308–312.

56. Q. Wang, Q. Sun, G. Chen, Y. Kawazoe & P. Jena, Vacancy-induced magnetism in ZnO thin films and nanowires, *Phys. Rev. B.,* 2008, **77,** 205411.

57. F. Beuneu & P. Monod, The Elliott relation in pnre metals, *Phys. Rev. B.,* 1978, **18,** 2422–2425.

58. F. Natali *et al,* Role of magnetic polarons in ferromagnetic GdN, *Phys. Rev. B.,* 2013, **87,** 035202.

59. D. Karmakar, *et al,* Ferromagnetism in Fe-doped ZnO nanocrystals: Experiment and theory, *Phys. Rev. B.,* 2007, **75,** 144404.

60. C. G. Jin *et al,* Tunable ferromagnetic behavior in Cr doped ZnO nanorod arrays through defect engineering, *J. Mater. Chem. C.,* 2014, **2,** 2992–2997.

61. L. Li *et al,* Ferromagnetism in polycrystalline Cr-doped ZnO films : Experiment and theory, *Solid State Commun.,* 2008, **146,** 420–424.

62. Q. Hou, Z. Xu, Z, X. Jia & C. Zhao, Effects of Ni doping and native point defects on magnetism of ZnO first-principles study, *J. Appl. Phys.,* 2018, **123,** 055106.

63. Z. Z. Weng, Z. G. Huang & W. X. Lin, Magnetism of Cr-doped ZnO with intrinsic defects, *J. Appl. Phys.,* 2012, **111,** 113915.

64. G. Weyer *et al,* Defect-related local magnetism at dilute Fe atoms in ion-implanted ZnO, *J. Appl. Phys.,* 2010, **102,** 113915.

65. V. N. Ivanovski *et al,* A study of defect structures in Fe-alloyed ZnO: Morphology, magnetism, and hyperfine interactions, *J. Appl. Phys.,* 2019, **126,** 125703.

66. B. Deng, Z. Guo & H. Sun, Theoretical study of Fe-doped -type ZnO, *Appl. Phys. Lett.,* 2010, **96,** 172106.

67. E.-Z. Liu & J. Z. Jiang, O-vacancy-mediated spin-spin interaction in Co-doped ZnO:

First-principles total-energy calculations, *J. Appl. Phys.*, 2010, **107,** 023909.
68. H. Liu *et al*, Structural, optical and magnetic properties of Cu and V co-doped ZnO nanoparticles, *Phys. E Low-dimensional Syst. Nanostructures.*,2013, **47,** 1–5.
69. K. Jug & V. A. Tikhomirov, Comparative Studies of Cation Doping of ZnO with Mn, Fe, and Co, *J. Phys. Chem. A.*, 2009, **113,** 11651–11655.
70. Y. Li *et al*, Micro-Brillouin scattering from a single isolated nanosphere, *Appl. Phys. Lett.*, 2006, **88,** 023112.
71. C. J. Cong, J. H. Hong & K. L. Zhang, Effect of atmosphere on the magnetic properties of the Co-doped ZnO magnetic semiconductors, *Mater. Chem. Phys.*, 2009, **113,** 435–440.
72. Z. Xiong *et al*, Oxygen enhanced ferromagnetism in Cr- doped ZnO films, *Appl. Phys. Lett.*, 2011, **99,** 052513.
73. Y. An, S. Wang, L. Duan, J. Liu & Z. Wu, Local Mn structure and room temperature ferromagnetism in Mn-doped $In_2O_3$ films, *Appl. Phys. Lett.*, 2013, **102,** 212411.